%% file: review-escape-reversion-260312-rev3.tex
\documentclass[12pt,notitlepage]{article}
\pdfoutput=1
\usepackage{graphicx,amssymb,amsfonts,amsmath,latexsym}
\usepackage[square,numbers]{natbib}%
\usepackage[bf,small]{caption}%
\usepackage{hyperref}

\input commands

\newcommand{\Vir}[1]{V(#1)}
\newcommand{\cell}[1]{{m_{#1}}}
\newcommand{\fit}[1]{{f_{#1}}}

\advance\voffset by -0.8in \advance\hoffset by -0.5in
\renewcommand{\cite}{\citep}

\newcommand{\esc}{\varepsilon}

\graphicspath{{.}{figs/}{figs-review/}}

\begin{document}
\title{Mathematical modeling of escape of HIV from cytotoxic T
  lymphocyte responses}

\author{\parbox{\textwidth}{\bc Vitaly V. Ganusov$^{1}$\thanks{Authors
      contributed equally to this report}, Richard A. Neher$^{2*}$ and
    Alan S.  Perelson$^3$ \ec}
  \\
  {\small $^1$Department of Microbiology, University of Tennessee,
    Knoxville, TN 37996, USA} \\
  {\small $^2$Max-Planck-Institute for Developmental Biology, 72070 T\"ubingen, Germany} \\
  {\small $^3$ Theoretical Biology and Biophysics, Los Alamos National
    Laboratory, MS K710}\\
  {\small  Los Alamos, 87545 NM, USA}\\
}
\maketitle

\begin{abstract}

  Human immunodeficiency virus (HIV-1 or simply HIV) induces a
  persistent infection, which in the absence of treatment leads to
  AIDS and death in almost all infected individuals.  HIV infection
  elicits a vigorous immune response starting about 2-3 weeks post
  infection that can lower the amount of virus in the body, but which
  cannot eradicate the virus. How HIV establishes a chronic infection
  in the face of a strong immune response remains poorly understood.
  It has been shown that HIV is able to rapidly change its proteins
  via mutation to evade recognition by virus-specific cytotoxic T
  lymphocytes (CTLs).  Typically, an HIV-infected patient will
  generate 4-12 CTL responses specific for parts of viral proteins
  called epitopes.  Such CTL responses lead to strong selective
  pressure to change the viral sequences encoding these epitopes so as
  to avoid CTL recognition. Indeed, the viral population ``escapes''
  from about half of the CTL responses by mutation in the first year.
  Here we review experimental data on HIV evolution in response to CTL
  pressure, mathematical models developed to explain this evolution,
  and highlight problems associated with the data and previous
  modeling efforts. We show that estimates of the strength of the
  epitope-specific CTL response depend on the method used to fit
  models to experimental data and on the assumptions made regarding
  how mutants are generated during infection. We illustrate that allowing
  CTL responses to decay over time may improve the fit to experimental
  data and provides higher estimates of the killing efficacy of
  HIV-specific CTLs. We also propose a novel method for simultaneously
  estimating the killing efficacy of multiple CTL populations specific
  for different epitopes of HIV using stochastic simulations. Lastly,
  we show that current estimates of the efficacy at which HIV-specific
  CTLs clear virus-infected cells can be improved by more frequent
  sampling of viral sequences and by combining data on sequence
  evolution with experimentally measured CTL dynamics.

 \vspace{0.5cm} {\bf Keywords}: acute HIV infection, escape
 mutations, CTL response, cost of escape, mathematical model.

 {\bf Abbreviations}: CTL, cytotoxic T lymphocyte, HIV, human
 immunodeficiency virus; SGA, single genome amplification.

 {\bf Short running title}: Escape in acute and chronic HIV infection

\end{abstract}

\section{Introduction}\label{sec:introduction}
Viruses replicate within cells. In order for the immune system to
recognize that a cell is infected, fragments of viral proteins or
peptides, typically 8-10 amino acids in length, called epitopes, are
presented on the surface of infected cells bound to major
histocompatibility complex (MHC) class I molecules
\citep{Yewdell.nri03,Neefjes.nri11}.  These complexes of viral
peptides and MHC-I molecules are then recognized by cytotoxic T
lymphocytes (CTLs) and this recognition leads to the death of
virus-infected cells \citep{Anthony.ir10}.

Because CTLs can recognize and kill virus-infected cells, they play an
important role in the control of many viral infections.  However, many
viruses, including cytomegalovirus and HIV persist, developing into
chronic infections despite very strong virus-specific CTL responses
\cite{vanLeeuwen.ir06,McMichael.nri10}. Viruses have evolved different
strategies to avoid recognition by CTL including downregulation of
MHC-I molecules \cite{Yewdell.ni02,Antoniou.i08} and generation of
mutants that are not recognized by CTLs, a process called ``escape''.
Some of these mutations affect binding of viral peptides to MHC-I
molecules and other mutations affect the ability of CTLs to recognize
the peptide-MHC complex \cite{Goulder.nri04}. Mutations at several
different sites within and sometimes outside the epitope sequence can
lead to viral escape
\cite{Allen.jv04,Draenert.jem04,Goulder.nri04,Prado.jv09}.  As a
result, viral mutants that are not recognized by epitope-specific CTLs
have a selective advantage and accumulate in the population over time
\cite{McMichael.nri10}.

Escape of HIV from CTL responses has been documented from months to
years after infection
\cite{Borrow.nm97,Jones.jem04,Goonetilleke.jem09,McMichael.nri10,%
  Allen.jv04,Draenert.jem04,Goulder.nri04} and escape from T cell
immunity may potentially drive disease progression \cite{Nowak.s91}.
Also escape from CTL responses may influence the efficacy of vaccines
that aim at stimulating T cell responses. Thus, understanding the
contribution of different factors to the rate and timing of viral
escape from CTL responses may help in designing better HIV vaccines.

A number of mathematical models have been developed to describe the
kinetics of viral escape from T cell immunity. Here we review some of
these models, show novel model developments, and discuss
directions of future research. 

\section{Modeling viral escape from a single CTL response}

During acute HIV infection there are several HIV-specific CTL
responses (on average around 7
\citep{Turnbull.ji09,Goonetilleke.jem09}), each recognizing a
different viral epitope. As the virus can escape from all these
responses escapes do not generally occur at the same time; some
escapes occur very early in infection and some late
\cite{Ganusov.jv11}.  Initial models of viral escape only examined
virus evolution in response to a single CTL response
\cite{Fernandez.jv05,Asquith.pb06,Ganusov.pcb06} and we will discuss
these first.

\subsection{Mathematical model}

To describe virus escape from a single CTL response we start with the
standard model for virus dynamics in which virus infects target cells,
i.e., cells susceptible to infection, and infected cells produce virus
(\fref{cartoon}). The model is formulated as a system of ordinary
differential equations

\beqa \Dt{T} &=& \lambda - dT - \beta T(V_w+V_m),
\label{eqn:T}\\
\Dt{I_w} &=& (1-\mu)\beta T V_w-(\delta+k) I_w, \label{eqn:Iw}\\
\Dt{I_m} &=& \mu \beta T V_w+ \beta T V_m-\delta I_m, \label{eqn:Im}\\
\Dt{V_w} &=& p_w I_w - c_V V_w, \label{eqn:Vw}\\
\Dt{V_m} &=& p_m I_m - c_V V_m, \label{eqn:Vm} \eea

\no where $T$ is the density of uninfected target cells, produced at
rate $\lambda$ and dying at per capita rate $d$. Infection is assumed
to occur via a mass-action like term with rate constant $\beta$. Cells
can be infected with either wild-type (the infecting strain) virus,
$V_w$, or escape mutant virus, $V_m$ leading to the generation of
infected cells, $I_w$ and $I_m$, respectively. Infected cells are
assumed to die at rate $\delta$ per cell due to viral cytopathogenic
effects, and at rate $k$ per cell due to killing by CTLs. Since the
escape variant is not recognized by CTLs, the term proportional to $k$
is absent in the $I_m$ equation. When an infecting virus is reverse
transcribed errors in copying occur at the mutation rate $\mu$. We
neglect back mutation from mutant to wild-type but this could easily
be added to the model. The constants $p_w$ and $p_m$ are the rates of
virus production by cells that are infected with the wild-type and
escape viruses, respectively, and $c_V$ is the clearance rate of free
viral particles.

\begin{figure}
\begin{center}
\includegraphics[width=0.5\textwidth]%
{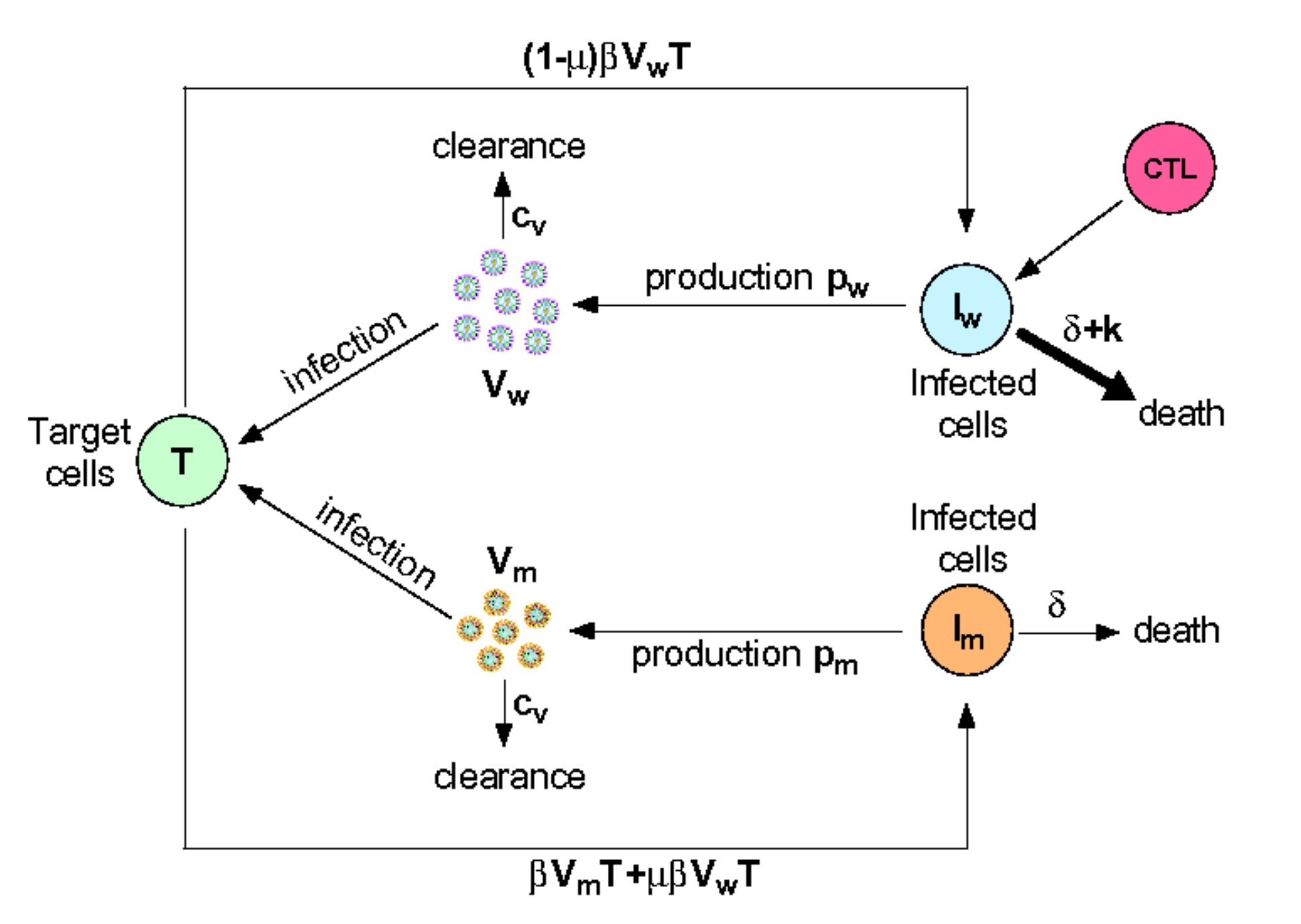}
\caption[cartoon]{Schematic illustration of the model of virus dynamics and escape
  from a CTL response. Symbols are defined in the text.}
  \label{fig:cartoon}
\end{center}
\end{figure}

In this model we made several simplifying assumptions.  We assumed
that the wild-type and escape viruses differ only in the rate of virus
production; generally $p_w\geq p_m$ (but see
\cite{Goonetilleke.jem09}). It is also possible that mutations that
lead to escape from the CTL response also affect viral infectivity,
$\beta$, especially if they occur in the envelope, reverse
transcriptase or integrase-coding regions of the viral genome. Because
viral particles are short-lived in vivo
\cite{Perelson.s96,Ramratnam.l99,Zhang.jv99}, a quasi-steady state is
rapidly established in which the density of viral particles is
proportional to the density of virus-infected cells, $V_w = I_w
p_w/c_V$ and $V_m = I_m p_m/c_V $. Then by substituting $r = p_w \beta
T/c_V $, $c=1-p_m/p_w$, $w=I_w$, $m=I_m$, we arrive at a simpler model
for the dynamics of the density of wild-type and mutant viruses:

\beqa %
\Dt{w(t)} & = & (1-\mu) r w(t) - (\dd+k) w(t), \label{eqn:wild.type}
\\
\Dt{m(t)} & = & r(1-c) m(t)+\mu r w(t) - \dd m(t), \label{eqn:mutant}
\eea

\no where $r$ and $r(1-c)$ are the replication rates of the wild type
and the mutant, respectively; $c$ is the cost of the escape mutation
defined as a selection coefficient \cite{Holland.jv91,Maree.jv00}. To
analyze this model, it is useful to rewrite
\erefs{wild.type}{mutant} in terms of the dynamics of the ratio of the
mutant to the wild-type density, $z(t)=m(t)/w(t)$

\beq \Dt{z(t)} = \Dt{m(t)}{1\over w(t)}- {z(t)\over w(t)}\Dt{z(t)} =
\mu r + z(t)\left(k-r(c-\mu)\right),
\label{eqn:main.ratio} \ee

\no Assuming a constant replication rate, $r$, and CTL killing rate, $k$,
\eref{main.ratio} can be solved analytically, where the ratio, $z(t)$,
increases exponentially with the rate $\esc = k-r(c-\mu)\approx k-cr$
(when $\mu\ll c$) which we call {\bf escape rate}, i.e.,

\beq z(t) = z_0 {\rm e}^{\esc t}+{\mu r\over \esc}\left({\rm e}^{\esc
    t} -1\right),
\label{eqn:z-mut2} \ee

\no where $z_0$ is the ratio at $t=0$. In the examples we give below,
the initial time, $t=0$, is the time when patients are first
identified as being HIV-infected and are enrolled in a clinical study.
This time of enrollment is likely to be several weeks after initial infection
\cite{Gasper-Smith.jv08,Gay.po11}. Similarly, the onset of CTL selection tends to be 
a few weeks after infection  \cite{Gasper-Smith.jv08,Gay.po11}. Equation
\ref{eqn:z-mut2} is only valid after the CTL selection has started and one has
to allow for the uncertainty of $t=0$ relative to the onset of selection by
adjusting $z_0$.

The dynamics of the ratio in \eref{z-mut2} is described by 3
parameters but only 2 parameters can in general be estimated from the
available viral sequence data \cite{Ganusov.pcb06,Ganusov.jv11}.
Therefore, two limiting cases of the general model can be found. If
rate of mutation is small and the escape variant is initially present
at a non-negligible frequency so that $z_0>0$, then the generation of
escape variants by mutation may be neglected, and the frequency of the
escape variant in the viral population is given by the logistic
equation

\beq f(t) = { f_0\over f_0 + (1-f_0){\rm e}^{-\esc
    t}},\label{eqn:frequency-logistic} \ee

\no where $f_0 = z_0/(1+z_0)$ is the initial frequency of the escape
variant in the population. Alternatively, if the initially escape
variant is not present and is generated by mutation from the wild-type
(i.e., $\mu>0$ and $z_0=0$) then the frequency of the escape variant
in the population is given by

\beq f(t) = {z(t)\over 1+z(t)} = {f_0\over f_0 + (1-f_0){\rm e}^{-\esc
    t}}\times\left(1-{\rm e}^{-\esc t}\right),\label{eqn:frequency}
\ee

\no where now $f_0 = \mu r/\esc$. It should be noted that at large
times ($t \esc \gg 1$), the dynamics predicted by \eref{frequency} and
\eref{frequency-logistic} are identical. In a later section we discuss
the situation where the escape variant is generated stochastically by
mutation. By fitting \eref{frequency} or \eref{frequency-logistic} to
experimental data, rates of viral escape $\esc$ from a given CTL
response can be estimated. In many previous studies, the logistic
equation (\eref{frequency-logistic}) that assumes that both wild-type
and escape variant were present at $t=0$, has been used
\cite{Asquith.pb06,Ganusov.pcb06,Goonetilleke.jem09,Ganusov.jv11}.

\begin{table}
\bc
\begin{tabular}{|p{0.3\textwidth}|l|l|l|}
  \hline
  quantity & symbol & value & references \\ \hline
  average mutation rate & $\mu$ & $2\times
  10^{-5}/\mbox{base}/\mbox{gen}$  & 
  \citep{Mansky.jv95}  \\
  net viral increase rate & $r-\delta$ & $0.9-1.3\ \mbox{day}^{-1}$ &  
  \citep{Ribeiro:2010p43150} \\
  free virus decay rate & $c_V$ & $23\ \mbox{day}^{-1}$  &
  \citep{Ramratnam.l99} \\
  infected cell death rate & $\delta$ & $1-2\ \mbox{day}^{-1}$  & 
  \citep{Perelson.s96,Bonhoeffer.tm03} \\
  virus production per cell (burst size) & $B$ & $5\times10^4$ & \cite{Chen.pnas07}\\
  effective population size & $N_e$ & $10^3-10^{7}$ & 
  \citep{Perelson.s96,Kouyos.tm06,Balagam.po11} \\ 
  virus infectivity & $\beta$ & varies & ---  \\
  \hline
\end{tabular}
\ec
\caption{Parameters determining the dynamics of HIV as estimated in previous studies. Here the viral increase rate is the rate at which
  HIV RNA accumulates in the blood during first weeks of 
  infection, $r=\beta\lambda p_w/(c_V d)$ (see \eref{Iw}). 
  There are no direct estimates of virus infectivity $\beta$ but
  its value can be adjusted to satisfy the condition $r-\delta\approx1$
  day$^{-1}$ observed during acute infection. Estimates of the effective 
  population size, which in the case of HIV infection
  is the number of virally infected cells, vary dramatically depending
  on the study. The rate of  
  virus production by infected cells is $p=N\delta$. 
  Not all virions produced by infected cells are
  infectious;  the ratio of infectious to noninfectious HIV
  is on the order of $10^{-2}-10^{-4}$ 
  \cite{Goto.av88,Ho.n95,Platt.jv10}.
}\label{tab:parameters-def}
\end{table}

The basic model assumes that the rate of viral escape from a given CTL
response is constant over time which in general implies a constant
rate of CTL-mediated killing of infected cells (determined by the
parameter $k$). Biologically, however, immune mediated selective
pressure is likely to change over time, for example, because of a 
change in the magnitude of the epitope-specific CD8+ T cell responses
\cite{Goonetilleke.jem09}. If the CTL killing efficacy, and as result
the escape rate, changes exponentially over time, e.g.,
$\esc(t)=\esc_0{\rm e}^{-at}$, the change in the frequency of the
mutant virus in the population over time can be obtained analytically
by solving \eref{main.ratio}

\beqa z(t) &=& \left( z_0{\rm e}^{\frac{\esc_0}{a}} + \frac{\mu}{a}
  \left[\phi\left(\frac{\esc_0}{a}\right)-\phi\left(\frac{\esc_0{\rm
          e}^{-at}}{a}\right)\right]\right)
\exp\left(-\frac{\esc_0{\rm e}^{-at}}{a}\right), \\
f(t) &=& \frac{z(t)}{1+z(t)},
\label{eqn:f-change}
\eea

\no where $\phi(x)= -\int_x^\infty \frac{{\rm e}^{-t}}{t}\id t$. As
before to reduce the number of parameters in the model we can assume
that either the escape variant is present at $t=0$ ($z_0>0$ and
$\mu=0$) or is generated by mutation ($z_0=0$ and $\mu>0$).

\subsection{Data and estimating model parameters}

Evasion of the CTL response by HIV occurs as the virus mutates
epitopes that are recognized by virus-specific CTLs. This escape
process can be studied by monitoring the sequence composition of the
viral population during infection. Over the last few years, detection
of viral escape mutations has been improved in two major ways.  First,
HIV RNA isolated from peripheral blood is diluted to the point that a
single RNA molecule is expected to be present in a given sample. Then
the RNA is reverse transcribed, amplified, and sequenced resulting in
the sequence for a given virus being obtained (so-called single genome
amplification and sequencing, SGA/S). When multiple viruses are
sequenced by SGA/S (in general about 10 to 20 per time point), the
sequences are compared at sites coding for a CTL epitope and changes
in the percent of the wild-type/transmitted sequence in the population
are followed over time \cite[\fref{escapes}A]{Goonetilleke.jem09}.
Second, deep sequencing can be done in which a relatively short RNA
region (about 150-300 nucleotides) is sequenced in the population
\cite{Fischer.po10}.  Although deep sequencing only allows one to
follow changes in a small region in the viral genome many more
sequences can be obtained than in the SGA/S protocol (from $10^2$ to
$10^4$).

An example, taken from ref. \citep{Goonetilleke.jem09}, of such time
course data of HIV immune escape is shown in \fref{escapes}. The
figure shows a schematic of the sequenced genomes and the frequencies
of escape mutations estimated as the fraction of times a mutation is
observed in the sample. Since samples are small (about 10-20 sequences
each), the frequency estimates come with substantial uncertainty.
Using data of this kind, we would like to infer escape rates
associated with CTL responses specific to different HIV epitopes using
the models discussed above.

Previously, data on viral escape have been analyzed by assuming that
some mutants are present at time $t=0$, using a logistic equation
\cite{Fernandez.jv05,Asquith.pb06,Ganusov.pcb06} and fitting the model
to the data using nonlinear least squares leaving $f_0$ completely
unconstrained. This yields results as shown in \fref{escapes}B.
Although this method often provides a reasonable description of the
data it does not weight the different data points according to the
uncertainty associated with them. Although weighted least squares can
take this uncertainty into account, we propose here to use a more
direct approach based on calculating the likelihood of the data given
our model. Similar methods have been developed in the context of
evolution experiments and the evolution of cancer
\citep{Illingworth.g11}. The likelihood of sampling a certain number
of mutants at different time points, given a particular escape rate,
$\esc$, and the initial mutant frequency, $f_0$, is derived as
follows.

Finding $k$ mutations in a sample of size $n$ when the true frequency
is $f(t)$ has the binomial probability
\begin{equation} {n \choose k} f(t)^k(1-f(t))^{n-k}.
  \label{eqn:binomial}
\end{equation}
The data set in general contains several samples of different sizes,
$n_i$, sampled at different times $t_i$. Given a frequency trajectory,
$f(t)$, such a data set therefore has the likelihood
\begin{equation}
  L = \prod_{i}{n_i \choose k_i} f(t_i)^{k_i}(1-f(t_i))^{n_i-k_i}. 
  \label{eqn:likelihood}
\end{equation}

Our model parameterizes the frequency trajectory of individual escapes
with the escape rate $\epsilon$, and the initial frequency $f_0$.
Ignoring all terms that do not depend on $\epsilon$ or $f_0$, we
obtain up to a constant

\begin{equation}
  {\cal L}=\log L =\sum_{i}\left[ k_i \ln(f(t_i))+(n_i-k_i)\ln 
    (1-f(t_i))\right].\label{eqn:logL}
\end{equation}

By maximizing this log-likelihood we obtain maximum likelihood
estimates of the parameters $f_0$ and $\epsilon$.  The confidence
interval of this estimator can be obtained by calculating the
curvature of the likelihood surface or by bootstrapping the data using
a binomial distribution \cite{Ganusov.jv11}. Furthermore, to constrain
some of the parameters of the model (e.g., the initial frequency of
the escape variant, see below) we can use a prior favoring some values
over other.

\begin{figure}
\begin{center}
A \hspace{0.4\textwidth}.\\
\includegraphics[width=0.33\columnwidth]%
{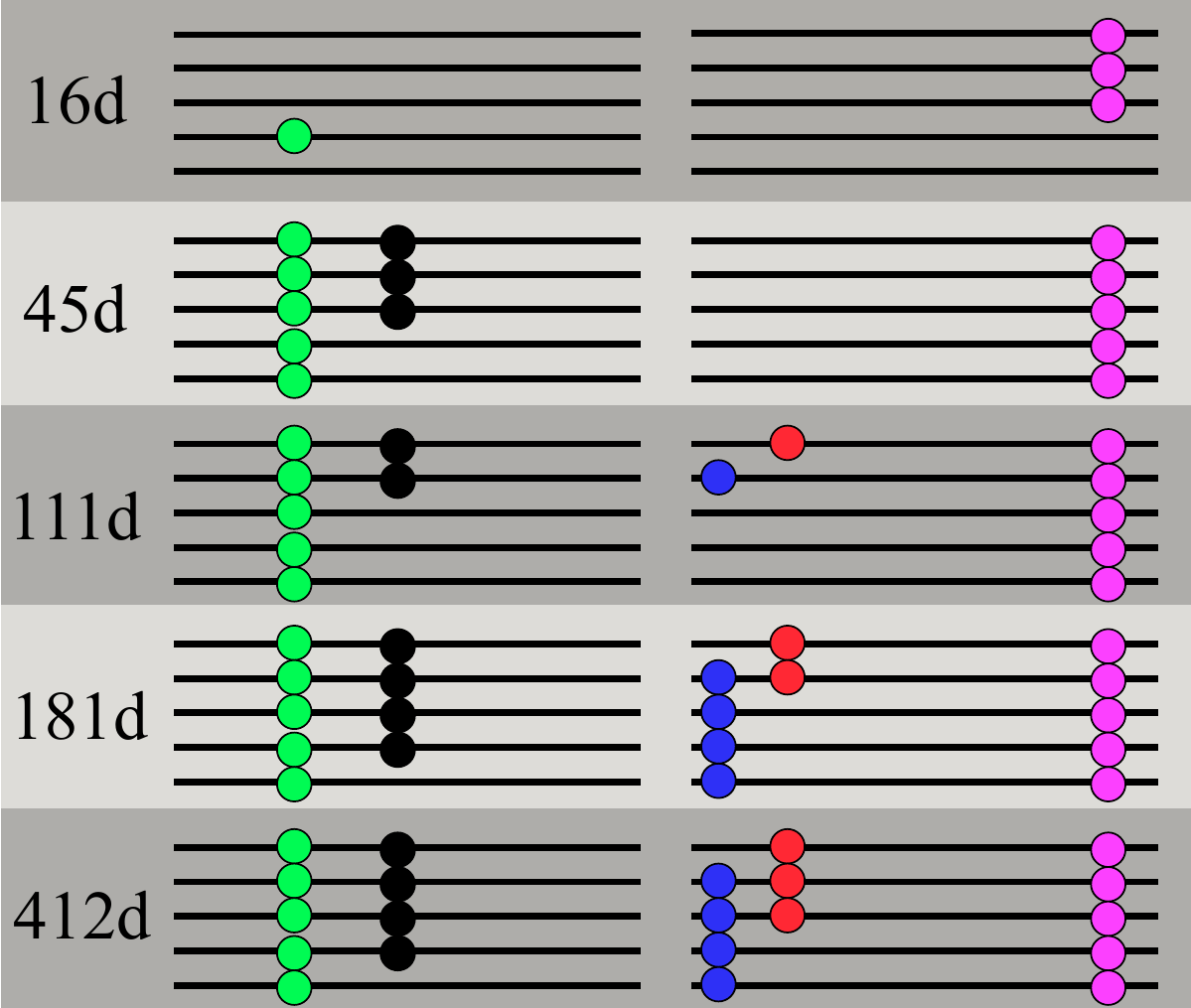}\\
B \hspace{0.4\textwidth}.\\
\includegraphics[width=0.45\columnwidth]%
{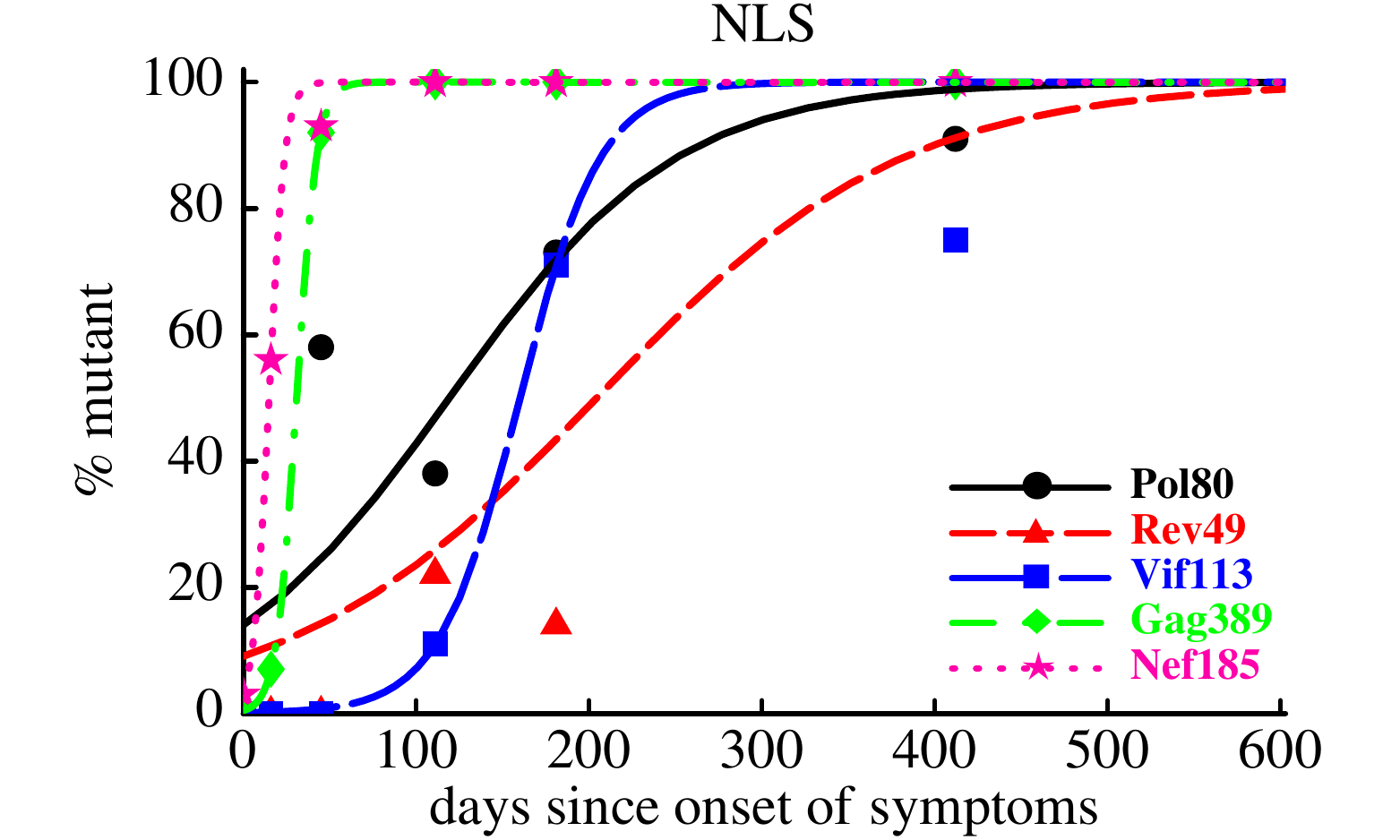}\\
C \hspace{0.4\textwidth}.\\
\includegraphics[width=0.45\columnwidth]%
{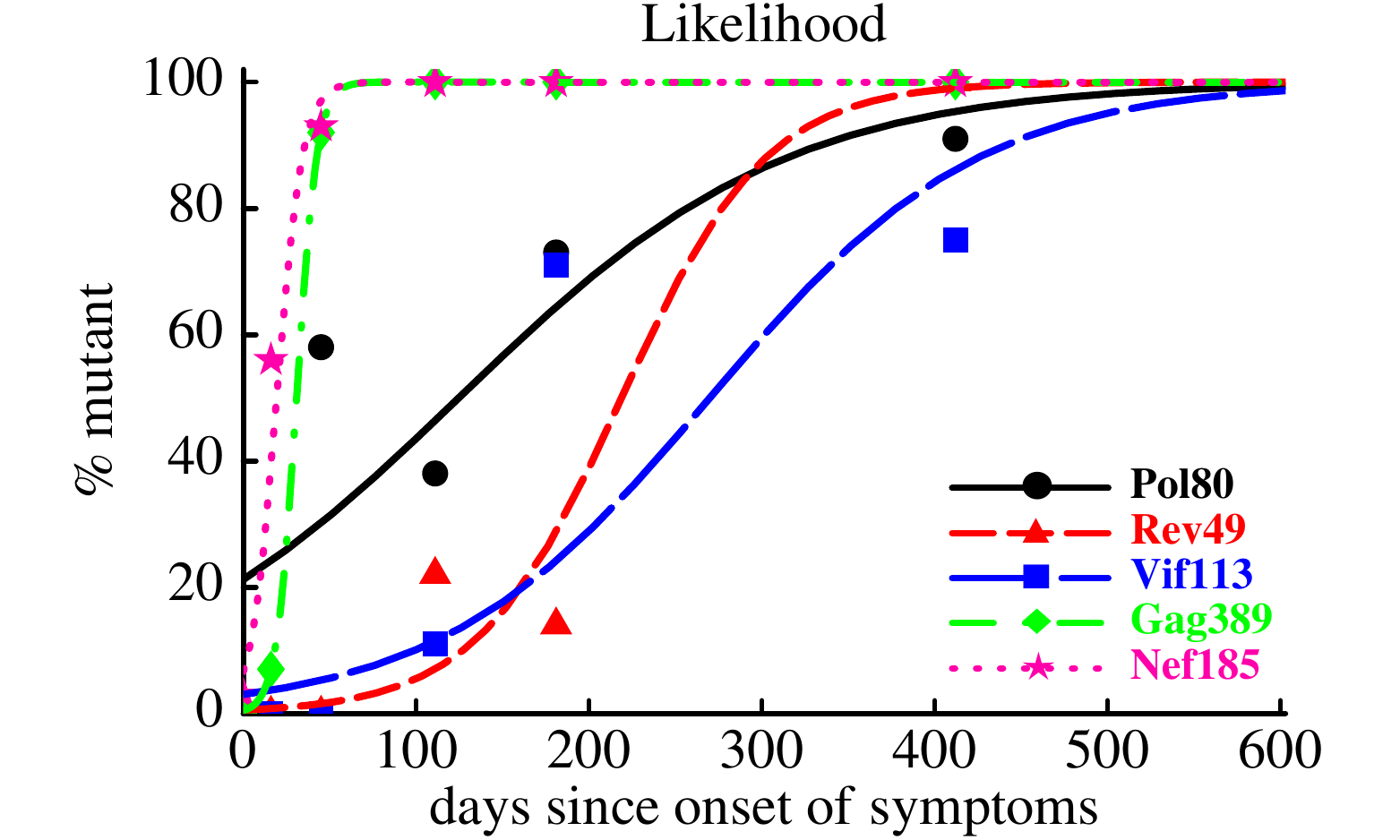}
\caption[likelihoo]{Schematic representation of experimental data on
  HIV escape from CTL responses (panel A) and fits of the mathematical
  model to such data (panels B and C).  In panel A, a small number of
  sequences covering either the 3' or 5' half of the HIV genome has
  been obtained at 5 different time points.  Escape mutations are
  indicated as colored dots. Typical sequence sample sizes range
  between 10 and 20.  In panels B and C we show the fits of the
  mathematical model (\eref{frequency-logistic}) to experimental data
  using nonlinear least squares (panel B) or likelihood (panel C)
  methods.  We estimate two parameters: the rate of escape, $\esc$,
  and the initial frequency of the escape variant in the population,
  $f_0$.  The estimated escape rate obtained by both methods is shown
  in \tref{parameters}.}
\label{fig:escapes}
\end{center}
\end{figure}

We applied both nonlinear least squares and likelihood methods to the
data shown in \fref{escapes}B on escape of HIV from five different CTL
responses in patient CH40 \cite{Goonetilleke.jem09,Ganusov.jv11}.
While both methods allow a reasonable description of the data, the
estimates of the escape rate from a given CTL response obtained by the
two methods are often different (\tref{parameters}).  For example, for
viral escape from the Rev49-specific CTL response, likelihood predicts
more rapid escape than the nonlinear least-squares method. In part,
this arises because of the oscillations in the measured frequency of
the mutant sequence in the viral population which initially increased,
then decreased, and then increased again. A similar argument applies
to the data on escape from the Pol80-specific CTL response.

\begin{table}
\bc
\begin{tabular}{|l|l|l|l|}
  \hline
  & \multicolumn{2}{c}{Single Epitope} & \multicolumn{1}{|c|}{Multiple
    Epitopes} \\ \cline{2-4} 
  Epitope & $\esc$, day$^{-1}$ (NLS) & $\esc$, day$^{-1}$ (Likelihood) &
  $\esc$, day$^{-1}$ (Likelihood) \\ 
  \hline
  Pol80&0.02&0.01& 0.05\\
  Rev49&0.01&0.02& 0.03\\
  Vif113&0.04&0.01& 0.02\\
  Gag389&0.17&0.17& 0.15\\
  Nef185&0.22&0.14& 0.18\\ \hline
\end{tabular}
\ec
\caption{Estimates of the rate at which HIV escapes from CTL responses
  specific to different viral epitopes. We fit a mathematical model 
  (\eref{frequency-logistic}) of
  HIV escape from a CTL response specific to a single viral epitope
  using nonlinear least squares (NLS) or
  maximum likelihood (\eref{logL}). The epitope is given in the first column and 
  the estimated escape rate, $\epsilon$, in the subsequent columns. 
  To investigate the influence of interference between escapes at 
  different epitopes we performed stochastic
  multi-locus simulations and determined the escape rates that maximize the 
  likelihood of observing the data averaged over several runs of the 
  stochastic simulation (see main text). 
  In these simulations, we assumed that CTL responses started
  30 days before the first patient sample was obtained; the estimated escape rates for
  early escapes are higher
  if this delay is shorter.
  The estimated rates are given in the column labeled ``multiple
  epitopes''.
  The estimated escape rates can depend 
  strongly on the model and method. One has to strike a 
  delicate balance between a too complicated model whose parameters
  cannot be determined due to insufficient data, and a too restrictive model
  with a well-defined optimal solution that is nevertheless inaccurate
  since the model was inappropriate. The rather flexible model  used in \cite{Ganusov.jv11}
  with two parameters per epitope results in large confidence
  intervals for estimated rates of viral escape (e.g., see \cite{Ganusov.jv11}).
}
\label{tab:parameters}
\end{table}

There are two problems with some of the model fits. First, some of the
fits predict a very high mutant frequency at time $t=0$ (e.g., for
Pol80 $f_0\approx 0.2$), which is inconsistent with the experimental
data. Second, the confidence intervals on the estimated escape rates
are very large (results not shown and \cite{Ganusov.jv11}).  The
underlying reason for the latter ambiguity is that without sufficient
data, the initial mutant frequency and escape rate are correlated, and
in general larger initial frequencies lead to lower escape rates.  To
reliably estimate two parameters, we have to have at least two
measurements where the mutant frequency is between 10 and 90\% (and
the data needs to be consistent with logistic growth; more on that
below).

\begin{figure}
\begin{center}
  \includegraphics[width=0.95\columnwidth]%
{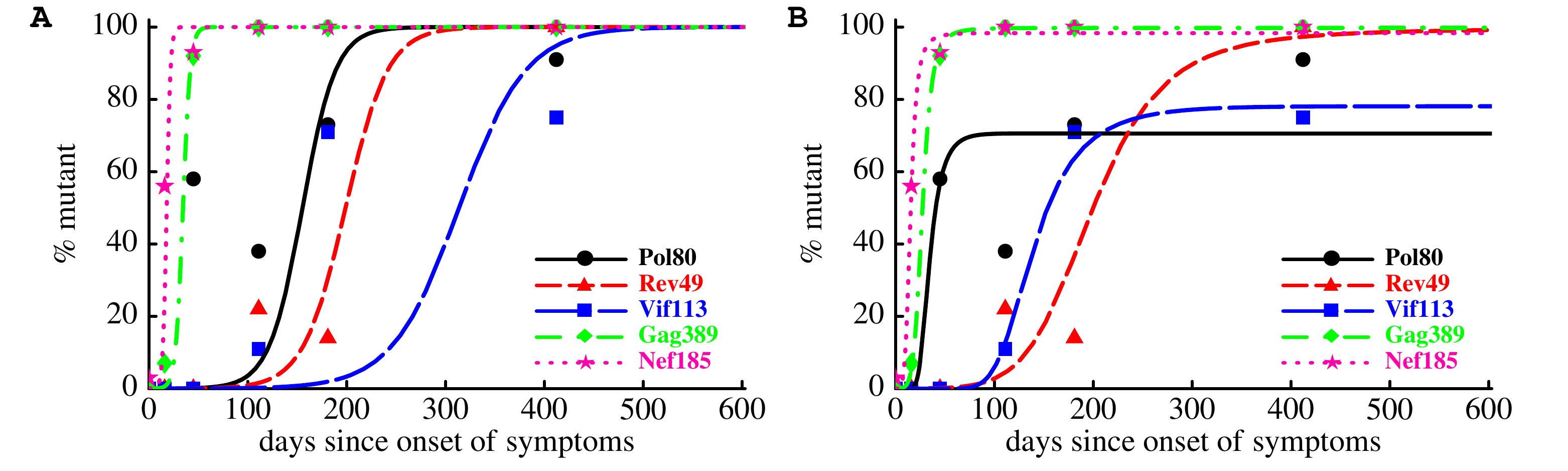}
  \caption[constrain]{Impact of constraining the initial mutant
    frequency, $f_0$, on the kinetics of viral escape. In panel A we
    show fits of the model (\eref{frequency-logistic}) to the sequence
    data obtained assuming that the initial frequency of escape
    variant is lower than $f_c=10^{-4}$. The fit is done by adding an
    extra penalizing term $10^{10^5(f_0-f_{c})}$ to the log-likelihood
    (\eref{logL}). The constraint leads to much higher estimates of
    the escape rate (\tref{constrain}) but in a poor description of
    the data.  In panel B in addition to the constraint to the initial
    mutant frequency we allow the escape rate to decline over the
    course of infection (\eref{f-change}). This extension improves the
    fit of the constrained model to data for 3 out 5 epitopes
    (\tref{constrain}).}
\label{fig:constrain}
\end{center}
\end{figure}

\begin{table}
\bc
\begin{tabular}{|l|l||l|l|}
  \hline
  Model & Constant $\esc$& \multicolumn{2}{|c|}{Decreasing $\esc$} \\ 
  \hline
  epitope & $\esc$, day$^{-1}$  & $\esc_0$ day$^{-1}$
  & $a$, day$^{-1}$\\ \hline
  Pol80$^*$ &0.06 &2.55 & 0.09\\
  Rev49     &0.05 &0.07 & 0.01\\
  Vif113$^*$&0.03 &0.76 & 0.02\\
  Gag389    &0.27  &0.53 & 0.04\\
  Nef185$^*$&0.51  &1.04 & 0.08\\ \hline
\end{tabular}
\ec
\caption{Estimates of the escape rate in the model where the initial
  frequency of the escape variant is constrained to be lower 
  than $f_c=10^{-4}$. In the 1$^{st}$ column we list the epitopes in
  which escape occurs. In the 2$^{nd}$ column we list estimates of the
  escape rate assuming a constant escape rate and using
  \eref{frequency-logistic} and \eref{logL} with a
  penalizing term $10^{10^5(f_0-f_{c})}$ added to the log
  likelihood. Model fits are shown in \fref{constrain}A. 
  In the 4$^{th}$ and 5$^{th}$ columns we list
  estimates of the initial escape rate and the rate of decline of the
  escape rate assuming that the escape rate declines over time using
  \eref{f-change} and \eref{logL}. Fits are shown in \fref{constrain}B. For the 3 
  epitopes indicated by $^*$ allowing the escape
  rate to change over time 
  significantly improved the quality of the
  model fit to data (likelihood ratio test, $p<0.0001$).
}\label{tab:constrain}
\end{table}

To circumvent both of these problems, one might be inclined to
constrain $f_0$ to be less than a prescribed cut-off (e.g.,
$f_0<10^{-4}$). Doing so reduces the variability of the fits and
generally results in larger estimates of the escape rates
(\fref{constrain}A and \tref{constrain}). At the same time, the fits
of the model to data on late escapes get substantially worse as early
data points are not described by the model. These inferior fits point
toward the inadequacy of the model.  One potential explanation for
this discrepancy between data and the model is that the escape rate
may be changing over the course of infection \cite{Ganusov.pcb06}.
Indeed, over time the magnitude of the CTL response may decrease
leading to a decreased selection pressure on the virus, and as a
result, a slower rate of escape later in infection. Indeed, allowing
the escape rate to change over the course of infection leads to a
significantly better description of the data at least for some escapes
(\fref{constrain}B and \tref{constrain}). Another feature that is
missing from the model is the simultaneous escape from multiple
epitopes, which we discuss at greater length below.

In summary, the model for viral escape from a single CTL response can
be used to estimate CTL-mediated pressure on the wild-type transmitted
virus using different statistical methods. If enough data is available
for a reliable estimate of $\esc$ and $f_0$ and the model predictions
are compatible with the observed data, direct estimation by fitting a
logistic involves the smallest number of assumptions. The estimated
escape rate might still be an underestimate due to variable selection
strength and the escape rate estimated using \eref{frequency} or
\eref{frequency-logistic} should be treated as the average escape rate
in the observed time period \cite{Ganusov.pcb06,Ganusov.jv11}. With
limited data, more robust estimates can be obtained by constraining
the initial frequency of escape mutants at the first time point, but
its validity rests on additional data on the time when the CTL
response to a given epitope is generated.

With these caveats in mind, the estimates nevertheless suggest that
virus-infected cells are killed by the virus-specific CTL responses
with rates ranging from 0.01 day$^{-1}$ to 0.4 day$^{-1}$
\cite{Asquith.pb06,Ganusov.jv11,Fischer.po10}, and if the escape rate
changes with the time since infection for a given epitope, killing
rates could be even higher (\tref{constrain}). Given that HIV-infected
cells have a death rate of $\sim1$ day$^{-1}$ \cite{Perelson.s96},
this work suggests that CTL responses contribute substantially to the
control of HIV at least during acute infection.

\subsection{Effects of sampling depth and frequency on fitting
  performance}

To perform a more systematic analysis of the fidelity of the different
fitting methods, we simulated escape trajectories using the
computational model for escape dynamics introduced below. From this
simulated data, we can produce a series of samples of different size
and mutant frequency and try to reconstruct the parameters that were
used in the simulation. The question we address here is: if we
want to improve estimates of the escape rate how should the data
collection be improved.

\fref{sample_size_depth} A\&B show two runs of the simulation with
shallow and infrequent  (A) and deep and frequent sampling (B). Deep
sampling will be readily achieved in forthcoming experiments since new
sequencing technologies allow deep sampling at low cost. The frequency
of sampling, however, will likely remain limited. Panel C shows how
well the escape rate of epitope 4 can be reconstructed from sample
series of different depth and frequency. The fitting procedure that
attempts to determine both $f_0$ and $\epsilon$ is rather noisy and
biased for small and infrequent sampling. Both deep and frequent
sampling allows one to overcome this problem. On the other hand, the
method that only fits the escape rate and assumes that variants are
present at a small frequency, $f_0$, when selection starts,
consistently underestimates the escape rate, but does not fluctuate a
lot. We will see below that this underestimate is a consequence of
delayed escape due to interference between different epitopes.

In order to estimate the escape rate and the initial frequency
reliably, we need to sample a trajectory at least twice at
intermediate frequency. This can be achieved both by deep or frequent
sampling. We would like to caution, however, that low frequencies are
very susceptible to fluctuations and rare variants found in a deep
sequencing experiment should not be assumed to follow a deterministic
trajectory.

\begin{figure}
\begin{center}
A \hspace{0.4\textwidth}.\\
\includegraphics[width=0.39\columnwidth,type=pdf,ext=.pdf,read=.pdf]%
{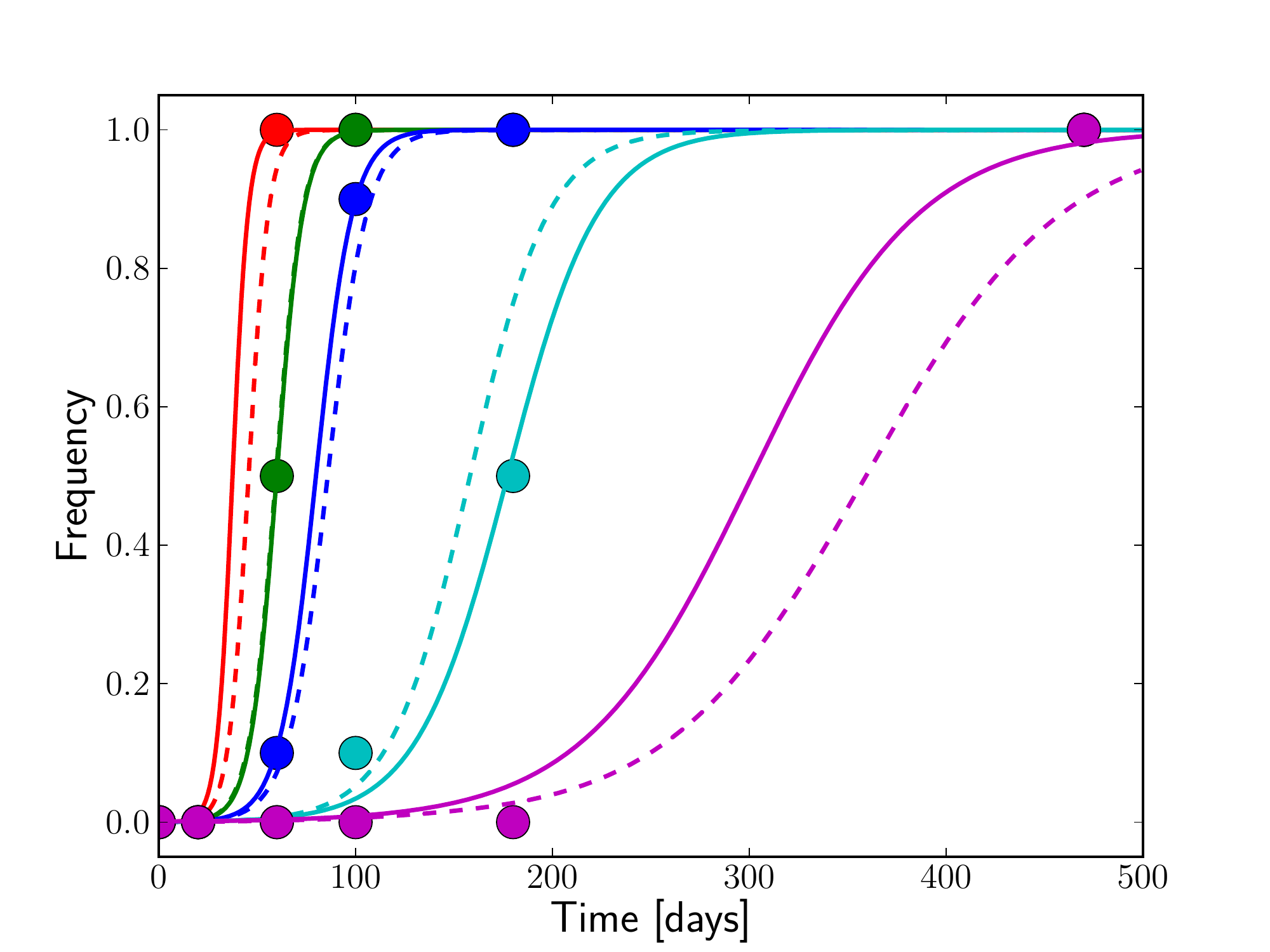}\\
B \hspace{0.4\textwidth}.\\
  \includegraphics[width=0.39\columnwidth,type=pdf,ext=.pdf,read=.pdf]%
{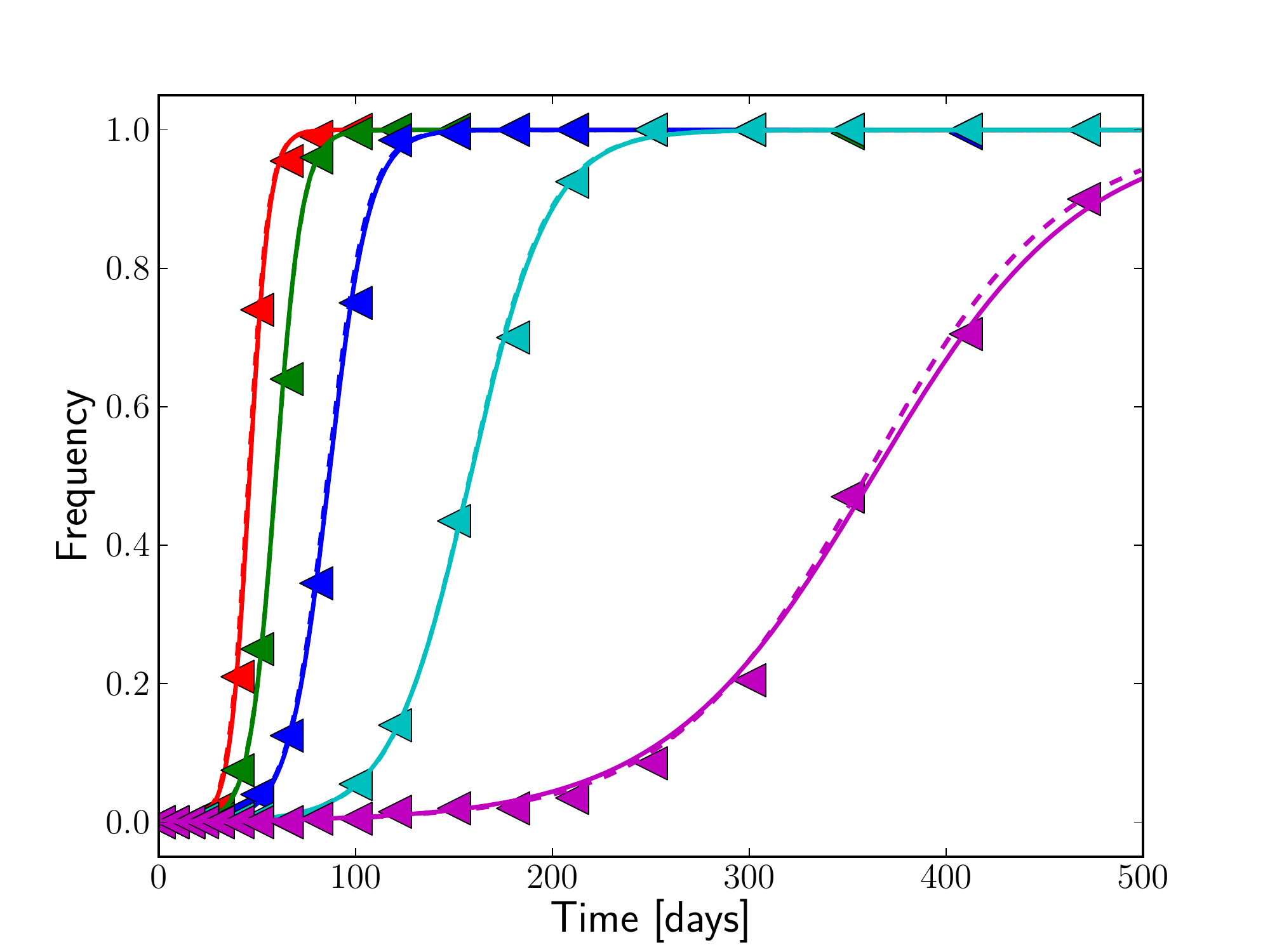}\\
C \hspace{0.4\textwidth}.\\
\includegraphics[width=0.39\columnwidth,type=pdf,ext=.pdf,read=.pdf]%
{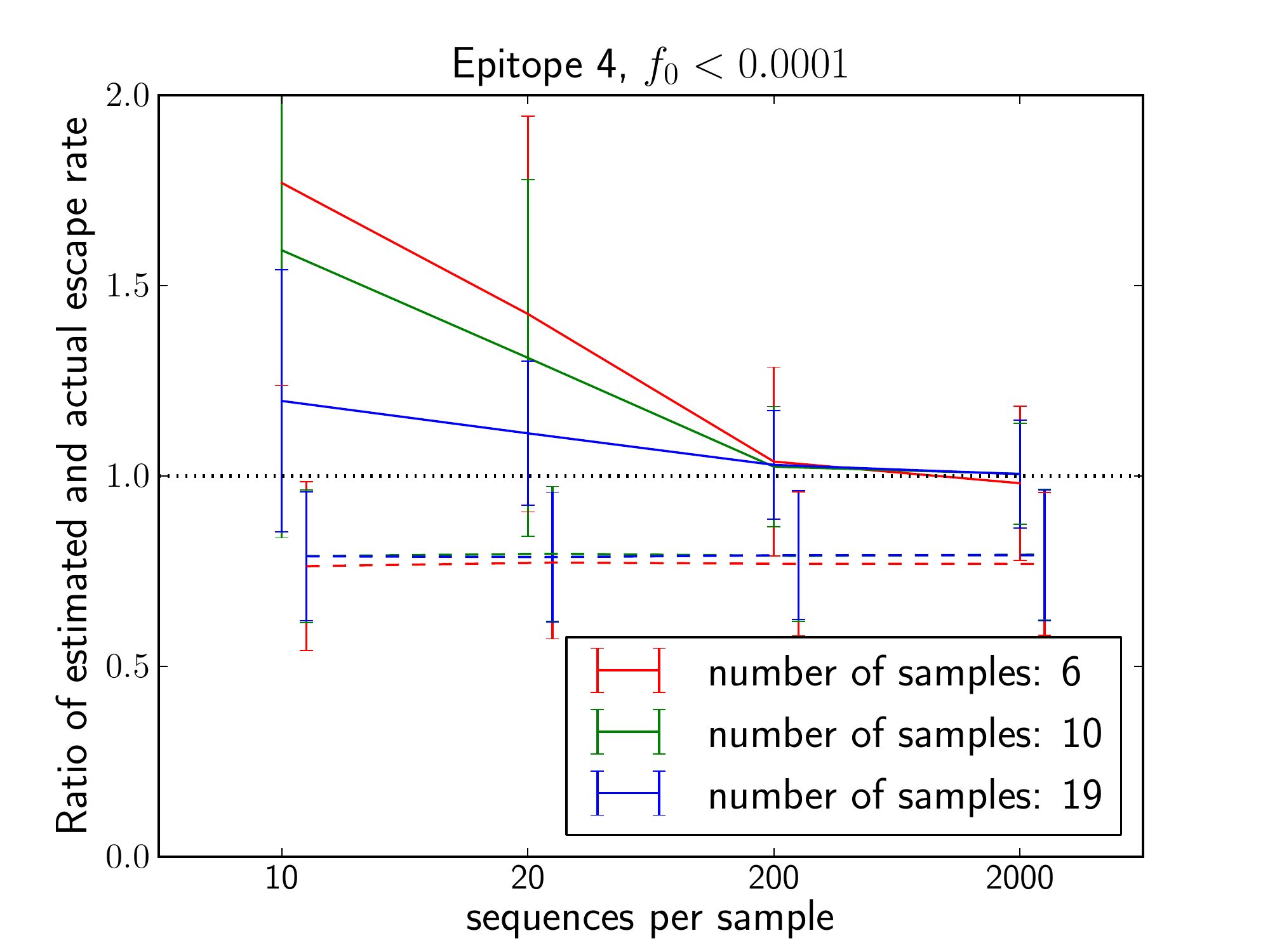}
\caption[labelInTOC]{Influence of sampling frequency and sampling
  depth on fidelity of esimates of the escape rate. Panels A and B
  show mutation frequencies in population samples for infrequent
  shallow sampling ($n=10$, A), and more frequent deep sampling
  ($n=200$, B). The actual mutation frequencies are shown as dashed
  lines, the sample frequencies are indicated by symbols, while the
  fitted trajectories as solid lines.  Obviously, more frequent and
  deeper sampling will improve the estimates of the escape rate. This
  is quantified in panel C. It shows the mean estimate of the escape
  rate of epitope 4 (left pointed triangles) and its standard
  deviation as a function of sampling depth for different sampling
  frequencies. The estimates are shown relative to the true value of
  the simulated escape rate, hence a systematic deviation from one
  represents a bias.  The dashed lines show the results of fitting
  only the escape rate, $\epsilon$, while fixing $f_0=10^{-4}$. Those
  fits show a systematic bias towards lower estimates, but have small
  variance and are insensitive to sample depth or frequency. The solid
  lines correspond to estimates where both $\epsilon$ and $f_0$ were
  fitted, while constraining $f_0$ to be smaller than $10^{-4}$.
  These fits show much larger variance and a strong bias at small
  sampling frequencies, but are unbiased at frequent and deep
  sampling. }
  \label{fig:sample_size_depth}
\end{center}
\end{figure}

\section{Modeling viral escape from multiple CTL responses}

While the model of viral escape from a single CTL response gives a
general idea of the rates involved in CTL escape, it is not a priori
obvious whether ignoring the simultaneous escape of other epitopes is
justified.  Different epitopes are encoded by the same viral genome
and as such are not independent. 
The analysis of multiple simultaneous CTL escapes is complicated by the large number of possible combinations of epitopes.  
 In the next section, we 
formulate a model for multiple simultaneous escapes as well as for
mutation and recombination that give rise to novel combinations of
epitopes.

\subsection{Mathematical model}

We assume that there are in total $n$ CTL responses that control viral
growth and, potentially, the virus can escape from all $n$ responses.
A CTL response that recognizes the $i^{th}$ epitope of the virus kills
virus-infected cells at rate $k_i$, and escaping from the $i^{th}$ CTL
response leads to a viral replicative fitness cost $c_i$.  We denote a
viral genome by a vector $\mathbf{i}=(i_1,i_2,\dots,i_n)$ with $i_j=0$
if there is no mutation in the $j^{th}$ CTL epitope and $i_j=1$ if
there is a mutation leading to escape from the $j^{th}$ CTL response.
The death rate of an escape variant due to the remaining CTL responses
is then simply $\sum_{j=1}^n k_j (1-i_j)$, where $k_1,k_2,\dots,k_n$
are the death rates of infected cells due to killing by the $j^{th}$
CTL response. Note that we have assumed that killing of infected cells
by different CTL responses is additive. Extending models for viral
escape with other mechanisms of CTL killing is an important area for
future research.

Escape from a given CTL response incurs a fitness cost to the virus.
Assuming multiplicative fitness, the fitness of a variant $\mathbf{i}$
is $\prod_j (1-c_j i_j)$.  Although there is evidence for compensatory
evolution in and around individual epitopes, we do not expect strong
epistasis between mutations in epitopes in different parts of the
genome.

Given that most HIV infections start with a single transmitted/founder virus
\cite{Keele.pnas08}, we need to describe the generation of the escape variants
from the founder strain. Even though the viral population during acute infection
may attain a large peak where there might be around $10^{10}$ infected cells, we
cannot assume that all possible viral genotypes are present early on. Because
$\mu^3\approx 10^{-14}$ is so small we do not expect to generate a virus with
more than two mutations in a single generation.
Multiple mutations therefore have to accumulate in the course of infection and
the appearance of these multiple mutants is delayed, as illustrated with
simulation data in \fref{escapes-stochastic-deterministic}.
Mutation dynamics therefore has to be included in the model.
Genotype ${\bf i}$ can arise by mutation with rate $\mu$ per epitope if a cell
gets infected with a viral strain lacking one of the mutations in ${\bf i}$:
\begin{equation}
  \mu \sum_{j\in {\bf i}} \Vir{{\bf i}\backslash j}
\end{equation}
where ${\bf i}\backslash j$ denotes genotype ${\bf i}$ without mutation $j$ and $\Vir{\bf i}$ is the abundance of virus with genotype ${\bf i}$.
We are mainly interested in the generation of escape mutations and will
therefore ignore back mutations. Similarly, we will for now ignore that 
genotypes are lost by mutations at all sites that have not yet escaped (this
term will be reinstantiated later).  Both of these contributions have negligible
effects on the dynamics since they do not involve genotypes that are favored by
selection. Furthermore, back mutations will occur at a slower rate because escape
mutation may occur at several positions in the epitope (8-10 amino acids) while
back mutations have to occur in the same place as the escape mutation.

\begin{figure}
\begin{center}
  \includegraphics[width=0.7\columnwidth]%
{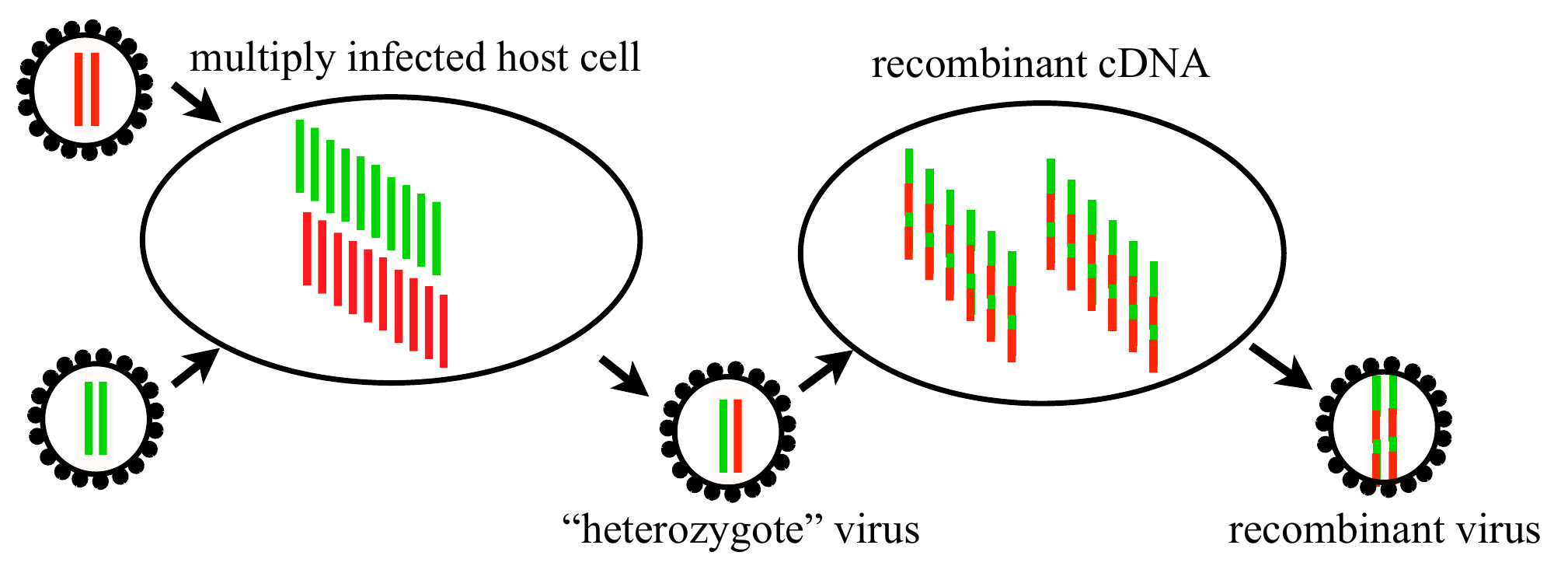}
  \caption[labelInTOC]{Each HIV particle contains two copies of its
    RNA genome, from which one complementary DNA strand is produced
    and integrated into the host cell genome. The two RNA strands are
    combined by template switching of the reverse transcriptase enzyme, which
    can happen up to 10 times per replication \citep{Levy:2004p23309}.
    The in vivo recombination rate, however, is limited by the
    probability that a host cell is infected by genetically distinct
    viruses, illustrated on the left. The effective recombination rate
    combining these two processes is estimated to be on the order of
    $10^{-5}$ per nucleotide per generation
    \citep{Neher:2010p32691,Batorsky:2011p40107,Josefsson:2011p42531},
    which implies a coinfection rate on the order of a few percent. }
  \label{fig:hivrec}
\end{center}
\end{figure}

In addition to mutation, novel genotypes can also be generated by
recombination of two existing HIV genomes.  Diversifying recombination
in HIV requires coinfection of a host cell with virions carrying
different genomes, which are crossed over by template switching in
subsequent generations \citep{Levy:2004p23309} (\fref{hivrec}). The
coinfection frequency was estimated to be on the order of a few
percent or less
\citep{Neher:2010p32691,Batorsky:2011p40107,Josefsson:2011p42531} and
is denoted here with the symbol $\rho$. We could extend the model to
include cells coinfected with different viral genotypes, but we will
simply assume that the fraction of viruses that are heterozygotes (see
\fref{hivrec}) with genotypes ${\bf j}$ and ${\bf k}$ is proportional
to product of the fraction of genotypes ${\bf j}$ and ${\bf k}$ in the
total population, i.e., $N^{-2}\Vir{{\bf j}} \Vir{{\bf k}}$, where
$N=\sum_{{\bf i}}\Vir{{\bf i}}$ is the total number of virus
particles. After infection with such a heterozygote virus, template
switching will produce a chimeric cDNA which is then integrated into
the target cell's genome.  Within this model, cells get infected with
the recombinant genotype ${\bf i}$ at rate
\begin{equation}
\label{eqn:recombinant_production}
\frac{\beta T \rho}{N}\sum_{{\bf j,k}} C({\bf i} | {\bf j,k}) \Vir{{\bf j}} \Vir{{\bf k}}
\end{equation}
where $C({\bf i} | {\bf j,k})$ is the probability of producing
genotype ${\bf i}$ from ${\bf j,k}$ by template switching. In this
expression, one factor of $N$ got canceled since
\eref{recombinant_production} accounts for the total production of
recombinant virus, rather than the fraction of total. The genotypes
that recombine are lost when producing the recombinant genotype, which
can be accounted for by a loss term $-\beta T \rho \Vir{\bf i}$.  The
mutation and recombination terms are easily incorporated into the
equations describing the viral population.  \beqa
\label{eqn:full_model}
\Dt{T} &=& d(T_0 - T) - \beta T\sum_{{\bf i}} \Vir{{\bf i}}, \label{eqn:TML}\\
\Dt{I({\bf i})} &=& \beta T \left(\Vir{{\bf i}}+ \mu \sum_{j\in {\bf i}} V({{\bf
i}\backslash j}) + \frac{\rho}{N}\sum_{{\bf j,k}} C({\bf i} | {\bf
j,k})\Vir{{\bf j}}\Vir{{\bf k}}  -\rho\Vir{{\bf i}}\right) \nonumber \\ &&- I({\bf
i})\left(\delta + \sum_{j=1}^n k_j (1-i_j)\right), \label{eqn:Im2} \\
\Dt{\Vir{{\bf i}}} &=& p({\bf i}) I({\bf i}) - c_V \Vir{{\bf i}}  
\label{eqn:Vm2} 
\eea 
where $I({\bf i})$ is the abundance of cells infected with strain ${\bf i}$.
The fitness costs of escape mutations are hidden in the rate of virus
production $p({\bf i})= p_0 \prod_j (1-c_j i_j)$.
Assuming the viral population is in a quasi-steady state, we substitute
$\Vir{{\bf i}} = \frac{p({\bf i}) I({\bf i})}{c_V}$, denote $\frac{\beta T p({\bf
i})}{c_V}$ by $f({\bf i})$, and normalize using $\cell{\bf i} = I({\bf i})/M$ with
$M=\sum_{\bf j} I({\bf j})$,  to obtain
\begin{equation}
\begin{split}
\Dt{}\cell{\bf i}(t) =& 
\left(f({\bf  i})(1-\rho) - \delta - \sum_{j=1}^n k_j (1-i_j)  -\frac{\dot M}{M}\right) \cell{\bf i} \\ 
&+\mu \sum_{j \in {\bf i}}  \fit{{\bf i}\backslash j}\cell{{\bf i} \backslash j}  +  \rho\frac{M}{N\beta T}\sum_{\bf j,k} C({\bf i | j,k})\fit{\bf j} \cell{\bf j}\fit{\bf k}\cell{\bf k},
\end{split}
\end{equation}
The term in big parentheses accounts for selection and the loss due to recombination, while the two terms on the second line account for the gain of genotype ${\bf i}$ through mutation and recombination, respectively. In the quasi-steady state, the average clearence of infected cells $\approx \delta M$ has to equal the number of new infections, given by the product of the number of virus particles $N$, the infectivity $\beta$, and the target cell number $T$. The prefactor of the recombination term is therefore approximately equal to $\rho/\delta$.
Since different viral genotypes reproduce with different efficiency $f({\bf i})$, the effective mutation and recombination rates at the level of infected cells have become genotype dependent. However, we will neglect this strain dependence in the following since it only leads to small changes in the mutational input and the recombination process. Defining the effective growth rate of a strain as $\epsilon_{\bf i} = \fit{\bf i} - \delta + \sum_{j=1}^n k_j (1-i_j)$ and the average growth rate $\langle \epsilon \rangle = \frac{\dot M}{M}$, and an effective recombination rate $\rho_e$, we can simplify the above to 
\begin{equation}
\label{eqn:multi_locus}
\Dt{}\cell{\bf i}(t) = \left(\epsilon_{\bf i}-\langle \epsilon \rangle \right) \cell{\bf i} +\mu \left(\sum_{j \in {\bf i}} \cell{\bf i \backslash j} - \sum_{j \notin {\bf i}}\cell{\bf i}\right) + \rho_{e}\left(\sum_{\bf j,k} C({\bf i | j,k}) \cell{\bf j}\cell{\bf k} - \cell{\bf i}\right),
\end{equation}
where we have restored the loss $\mu\sum_{j \notin {\bf i}}\cell{\bf i}$ due to mutations at wild-type epitopes. 

The three terms account for changes in frequency due to differential
replication and killing, mutation, and recombination, respectively.
The mutation and recombination terms account both for influx and
efflux of genotypes. The effective recombination rate should be thought of as the rate at which novel genotypes are produced from existing genotypes and accounts for coinfection, copackaging, and the average relatedness of copacked genomes.
Within our additive model, the growth rate
$\epsilon_{{\bf i}}$ is a sum of terms accounting for the fitness
costs of the escape mutations and the avoided killing. 

Equation~\ref{eqn:multi_locus} provides a simpler description of the
viral population than \erefs{full_model}{Vm2}.  The dynamics of the
free virus has been slaved to the frequencies of infected cells and
the complex parameters describing virus reproduction and killing have
been subsumed in a simple growth rate. Models of this type have been
studied intensively in population genetics. For a review of
theoretical work on the evolution of multi-locus systems we refer the
reader to \citep{Neher:2011p45096}. \citet{daSilva.g12} has introduced
a similar model and investigated how different assumptions about
mutation rates, coinfection probability, and CTL killing efficacy
influence the number and timing of escapes.

Equation~\ref{eqn:multi_locus} still describes deterministic dynamics.
Stochastic effects, however, are important whenever a particular
genotype is present in small numbers. The stochastic features of the
dynamics can be easily incorporated in computer simulations where each
individual can replicate, mutate, and recombine with a certain
probability each time step, see below. Examples of such stochastic
simulations are shown in \fref{escapes-stochastic-deterministic},
where the frequencies of escape mutations in stochastic simulations
are compared to the deterministic solution of the system.  The
stochastic trajectories deviate significantly from the deterministic
ones, in particular, when the recombination rate is low.

To appreciate how stochasticity, in combination with selection, and recombination can
affect the viral population dynamics, it is useful to consider the
extreme case of no recombination, i.e., ~asexual evolution. To
produce a genotype with multiple beneficial mutations, a series of
mutations in the same lineage is required since mutations
happening on different genomes cannot be combined in the absence of
recombination. Hence the only mutations that can successfully spread
through the population are those that happen on already very fit virus
and produce new exceptionally fit genomes. All other mutations, even
if beneficial, are lost since they are outcompeted by fitter genotypes
-- a phenomenon often called selective interference \citep{Gerrish:1998p5933}. Since this
seeding of new exceptionally fit genotypes is a rare process that
involves a very small number of viruses, and the existence or absence
of such fit virus determines the future dynamics, the stochasticity of
the population dynamics is important.

\begin{figure}
\begin{center}
  \includegraphics[width=0.48\columnwidth,type=pdf,ext=.pdf,read=.pdf]%
{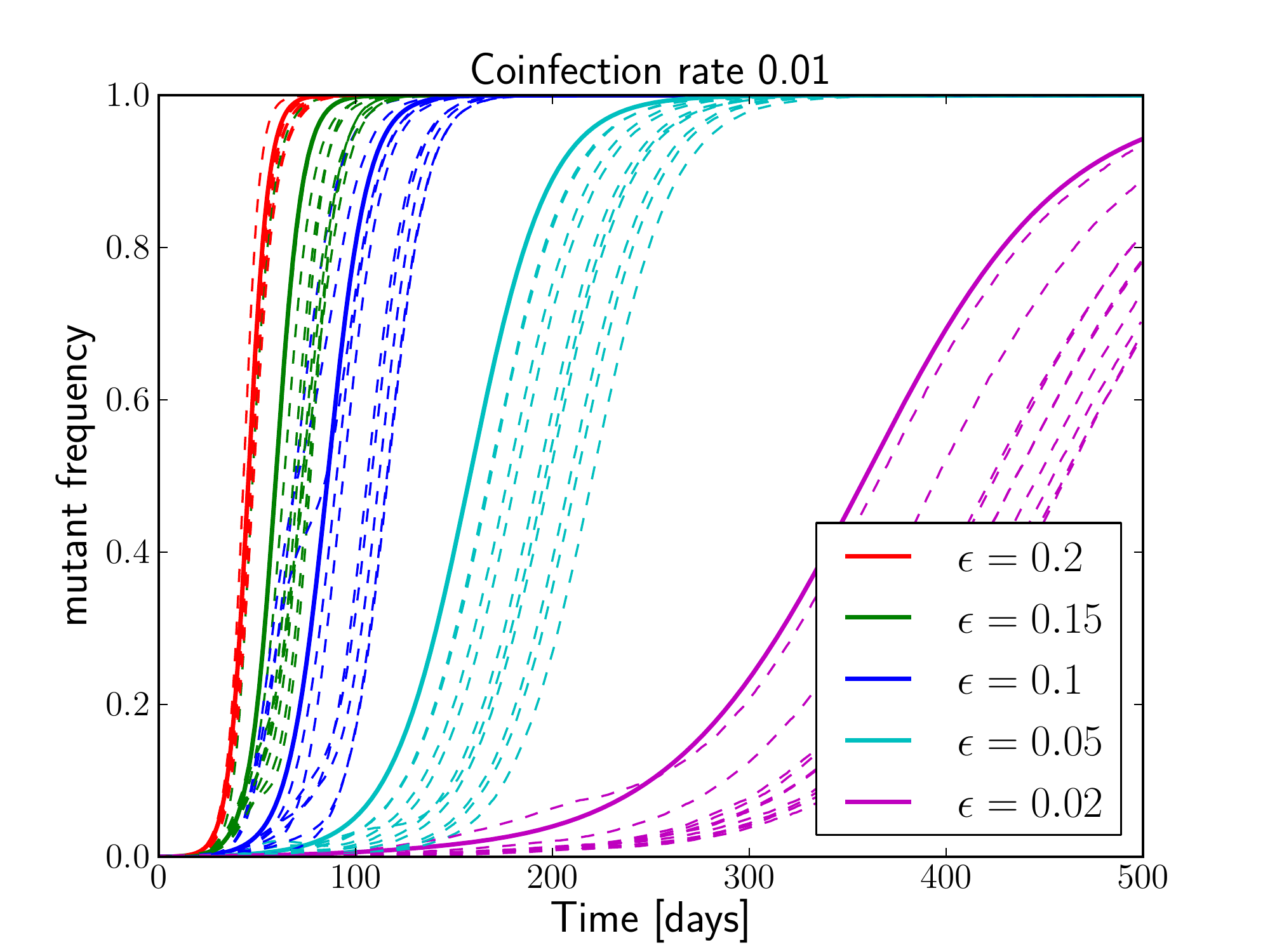}
  \includegraphics[width=0.48\columnwidth,type=pdf,ext=.pdf,read=.pdf]%
{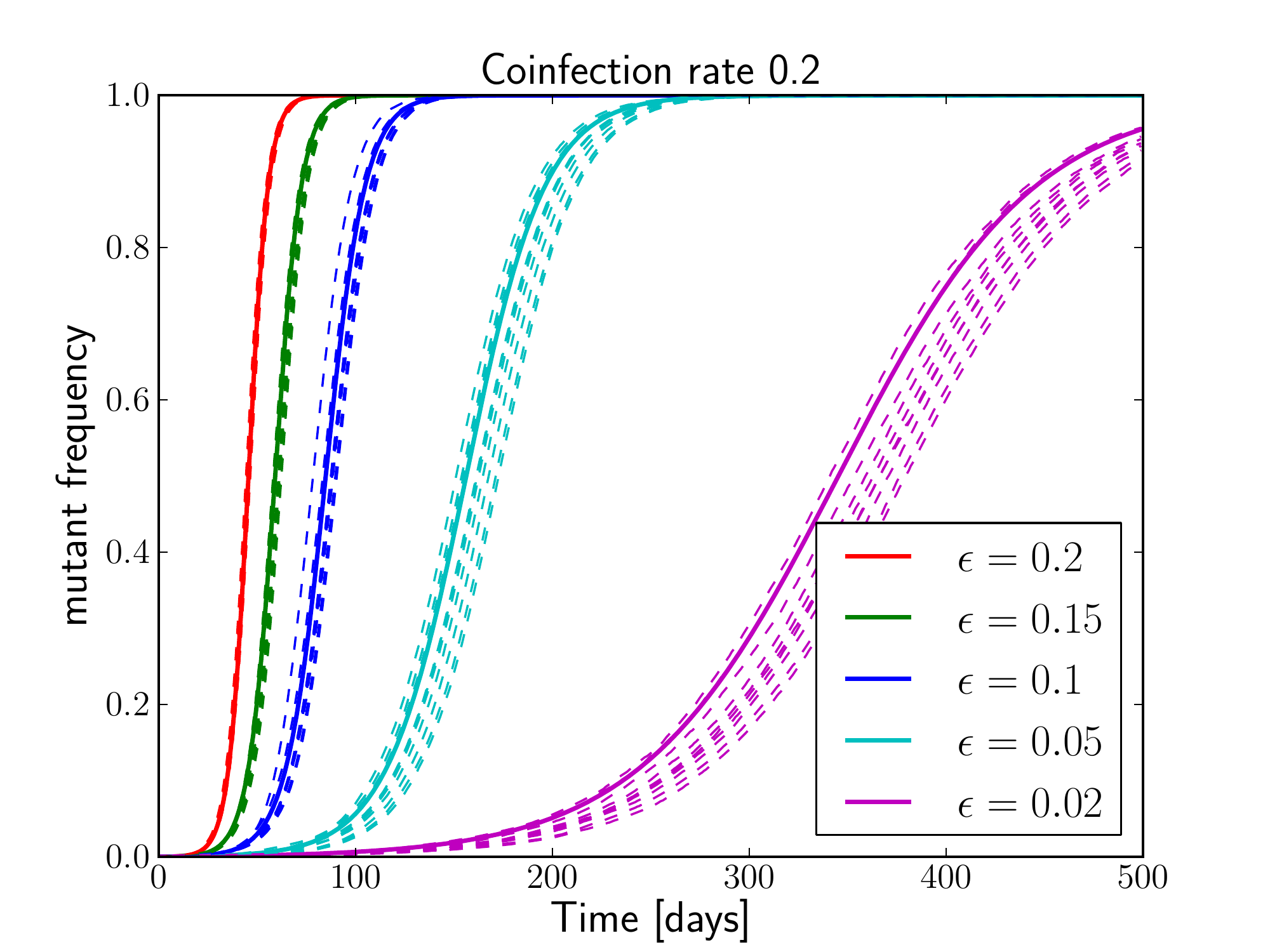}
  \caption[escapes2]{Influence of stochastic effects on CTL escapes.
    The rise of escape mutations in the stochastic model (dashed
    lines, 10 realizations) is delayed relative to the deterministic
    model (solid lines) at low coinfection rates (left panel, coinfection
    rate 0.01). This delay is much shorter at increased coinfection
    frequencies (right panel, coinfection rate 0.2), suggesting that
    the delay is mainly due to interference between epitopes.  The population size is
    $10^6$.  }
   \label{fig:escapes-stochastic-deterministic}
\end{center}
\end{figure}

A particular escape mutation might have to arise multiple times until it is
finally falls onto a genome that is successful. This interference can
substantially delay the accumulation of mutations as is apparent in
\fref{escapes-stochastic-deterministic}, which shows that competition between
different mutations can have substantial effects on the allele frequency
trajectories. When such delays are not accounted for, the estimates of escape
rates can be biased as apparent in \fref{sample_size_depth}.

At large recombination rates genotypes are constantly taken apart and
reassembled from the existing genetic variation. Escape mutations that
happen on different genomes can be combined by recombination to
produce better adapted virus. Hence recombination accelerates the
production of recombinant virus and reduces the fluctuations of allele
frequency trajectories.

The crossover between a more or less asexual population to one that
behaves like a fully sexual one depends on the strength of selection.
Selection operates on the fitness of entire genotypes and changes the
genetic composition of the population on time scales that are
inversely proportional to the fitness differences in the population.
If this time scale is much shorter than the inverse recombination rate,
recombination has a small impact on the dynamics. It does, however,
occasionally produce new genotypes similar to mutation.  If
recombination is faster than selection, genotypes are taken apart and
reassembled by recombination before their frequency is changed
substantially by selection. In this case the frequency of the genotype
is the product of the frequencies of the alleles it is composed of. In
other words, recombination decouples different loci along the genome
and the dynamics of allele frequencies at each locus are well
described by the single epitope model.

The recombination rate of HIV is such that both of these limits are
important in different phases of the infection.  The frequency of
recombination between distant parts on the viral genome (distance
$l>1$ kb) is limited by the probability of coinfection, which is
estimated to be on the order of a few percent or less
\citep{Neher:2010p32691,Batorsky:2011p40107,Josefsson:2011p42531}. For
loci closer together than a distance $l$, the recombination rate will be
approximately $10^{-5}\times l$ per generation
\citep{Neher:2010p32691}. The parameters estimated above suggest that
changes in genotype frequencies are much more rapid than decoupling by
recombination, at least during the early part of the infection. Hence
in order to estimate the parameters of the model, we have to take the
complex dynamics of a stochastically evolving population into account.
During later stages of the infection, changes in genotype frequencies
are much less rapid, such that distant parts of the viral genome are
essentially decoupled. The effect of selection in partly sexual
populations like HIV has been studied in greater detail in
\citep{Rouzine:2005p17398,Neher:2010p30641,Neher:2011p42539}.

In essence, the two regimes of high and low recombination differ in
what the relevant dynamical variables are. In the early regime where
selection is strong, fit viral strains are amplified by selection,
while mutation and recombination produce novel strains at a smaller
rate. The relevant quantities are the frequencies of different
strains, which happen to be the variables of our model.
Later in infection, however, when recombination dominates over
selection, the frequencies of mutations evolve
approximately independently of each other and genotypes frequencies
are slaved to these mutation frequencies \citep{Neher:2011p45096}.

Whether one or the other description is appropriate matters for the
interpretation of the data. The rapid rise of several mutations that
occur together as one genotype is most likely driven by the joint
effect of all of these mutations. Estimates of an escape rate from the
slope of the frequency trajectory would therefore correspond to an
escape rate of a genotype rather than an individual mutation. For
example, escape of HIV from Gag389- and Nef185-specific CTL responses
occurs within the same time frame and therefore, our estimates of viral
escape from individual responses (0.17 day$^{-1}$ and 0.14 day$^{-1}$,
respectively) likely represent simultaneous escape from both responses
($\approx$ 0.16 day$^{-1}$, see \tref{parameters} and
\fref{escapes}C).

Later in the infection, when recombination and selection are of
comparable strength, the trajectory of a particular mutations would
reflect selection on this mutation alone, even if other mutations
escape at the same time.

The problem of the accumulation of competing beneficial mutation in
large sexual and asexual population is an active area of research in
population genetics
\citep{Rouzine:2003p33590,Desai:2007p954,Rouzine:2005p17398,Neher:2010p30641}.
Analytic results have only been obtained for drastically simplified
models, which are not suitable for the inference of model parameters
of the sort we are interested in here. On the other hand, we are
typically interested in the evolution of just a few sites, which can
be efficiently simulated.

\subsection{Simulation of multiple CTL escapes}

We have implemented the simplified model described above as a computer
simulation using a discrete time evolution scheme. The simulation
keeps track of the abundance $\cell{\bf i}$ of each of the $2^n$  possible viral
genotypes, where $n$ is the number of epitopes. In each generation, $\cell{\bf i}$ is
replaced by with $\cell{\bf i}e^{\esc({\bf i}) - \langle \esc\rangle}$,
which accounts for selection. To implement recombination, we calculate
the distribution of recombinant genomes resulting from random pairing
of genotypes after selection. It is assumed that all loci reassorted
at random, which is justified if all epitopes are further apart than
1000 bp. A genetic map could be implemented easily. A fraction, $\rho$,
of the population is replaced by recombinant genomes in each generation. Similarly,
mutations change the genotype distribution by moving $\mu \cell{{\bf
    i}\backslash j}$ individuals with genotype ${\bf i}\backslash j$ to genotype
$\cell{\bf i}$ and vice versa for every possible ${\bf i}$ and $j$.
To account for the stochastic nature of viral reproduction, the
population is resampled according to a Poisson distribution after
selection, recombination, and mutation. The average population size
can be set at will in this resampling step.  The program source code
and brief documentation is available as supplementary information.  Due to
recombination, the computational complexity scales as $3^n$ and a
simulation of $n=10$ epitopes for 500 days runs for about one second
on a typical 2011 desktop computer. The simulation is built using the 
a general library FFPopSim for multi-locus evolution. 
The source code, documentation, and a python wrapper are 
available from \url{http://code.google.com/p/ffpopsim}.

\subsubsection{Inferring escape rates by multi-locus simulations}
Given our model of multi-epitope viral escape and a simulation to generate 
trajectories, we can try to infer the escape rates by adjusting the parameters
of the model to maximize the likelihood of the observed escape trajectories. 
In absence of any tested fitting
procedure for such a problem, we simulated the dynamics for a large
number of parameters and determined the likelihood of sampling the
observed mutations from the simulation (we tested 21 values of the
escape rates for each epitope, i.e., $21^5$ rate combinations).
 
In addition to the escape rates of the different epitopes, we
introduced an additional parameter $\tau$ that specifies the onset of
CTL selection relative to the time of the first available patient sample. 
Other parameters such as $\mu=2\times 10^{-5}$ and $\rho=0.01$ are taken from the literature. The
population is initialized as a homogeneous population without any escape
mutations $\tau$ generations prior to the first sample.

The likelihood of the data given the escape mutant frequencies is calculated 
using \eref{likelihood}.
Empirically, we find that there is a single (broad) maximum of the
likelihood surface and that fits are best with CTL selection onset
$20-30$ days before the first sample. The values in \tref{parameters} correspond to $\tau=-30$.
However, we would like to emphasize that the agreement between
the simulation and the data is never terribly
good, which, as discussed above, is possibly due to changing selection
pressure over time.

\section{Conclusions \& Future directions}

We have discussed several models of the dynamics of immune escape at
single or multiple loci. We have shown how the model fit depends on
the assumptions made by the model. By applying the inference
procedures to simulated data, we investigated how the sampling depth
and sampling frequency affects the fidelity of the estimates.

The models and procedures outlined have a number of short-comings
that need to be addressed to obtain more meaningful estimates of the
parameters governing the co-evolution of the viral population and the
immune system. The models are both too simple and too complex.  On
one hand, there is mounting evidence that the models miss several
important aspects of the immune system/virus interaction. On the other
hand, the models already contain too many parameters to allow for their
robust estimation from the available data.

It has recently become clear that the adaptive immune system is able
to control the virus by other means than the direct killing of
infected cells, for example, by production of antiviral cytokines and
chemokines \citep{Ganusov.jv11,Klatt10,Wong10}.  Furthermore, the
immune systems produces a very dynamic environment for the virus where
the selection pressure on different epitopes is changing. We have
generalized the single locus models to allow for exponentially
decaying escape rates, but introducing one additional parameter per
locus makes the fit near degenerate unless a constraint on the initial
frequency of the escape mutant is introduced.  We have also ignored
the possibility of compensatory mutations, competition between multiple
escape variants at a single epitope, and epistatic interactions between
mutations.

Another potential extension of the model is to allow the processes of
mutation and selection due to escape from CTL responses to start at
different times post infection. Indeed, mutation from the founder
virus starts at the beginning of infection while most CTL
responses do not arise until 2-4 weeks post infection
\cite{McMichael.nri10}. Also, it is not well understood how multiple CTLs that
are specific for different viral epitopes interact to kill virally infected cells, 
e.g., whether the death rate of cells expressing different viral
epitopes is the sum of the killing rates due to individual
epitope-specific CTL responses. Recent work has shown that competition
between different CTL responses may influence the timing and speed of
viral escape \cite{Ganusov.jv11}.

The analysis of multi-locus data is hampered by the large number of
possible genotypes, which grows exponentially with the number of loci
considered. The dynamics of this genotype distribution is governed by
a non-linear equation and solving the model involves considerable
computational effort, such that one would expect fitting parameters of
the model to be slow and ridden with many suboptimal local minima. The
problem, however, is not as daunting as it seems.

The majority of the possible genotypes will never exist and the
population is always dominated by a small number genotypes. Furthermore, the escape
mutations accumulate in the inverse order of their escape rates, which
implies that early mutations affect the dynamics of later mutations,
but not vice versa. 

Lacking an analytical solution of the multi-locus dynamics, fitting
parameters will require repeated simulation of the population dynamics
and comparison of the simulated trajectories with the data. The
underlying dynamics of the population, however, is stochastic and
different runs of a stochastic simulation will result in
different outcomes, such that fitting to a stochastic simulation is
ambiguous.

All of these additions will provide interesting future directions, 
particularly when deep and dense data are available to constrain the
models.

\section{Acknowledgements}
This work began with discussions between ASP and RN at a Kavli
Institute of Theoretical Physics workshop supported by NSF grant
PHY05-51164. This work was performed under the auspices of the U.S.
Department of Energy under contract DE-AC52-06NA25396, and supported
by NIH grant R37-AI028433 and the National Center for Research
Resources and the Office of Research Infrastructure Programs (ORIP)
through grant 8R01-OD011095-21 (ASP). RAN is supported by ERC Starting
Grant no.~260686(HIVEVO).

\bibliography{escapes}

\end{document}

%% file: commands.tex
\newcommand{\bba}[1]{\begin{alertblock}{#1}}
\newcommand{\eba}{\end{alertblock}}
\newcommand{\bbe}[1]{\begin{exampleblock}{#1}}
\newcommand{\ebe}{\end{exampleblock}}

\newcommand{\bii}{\begin{itemize}}
\newcommand{\eii}{\end{itemize}}
\newcommand{\bie}{\begin{enumerate}}
\newcommand{\eie}{\end{enumerate}}
\newcommand{\beq}{\begin{equation}}
\newcommand{\ee}{\end{equation}}
\newcommand{\eeq}{\end{equation}}
\newcommand{\beqa}{\begin{eqnarray}}
\newcommand{\eea}{\end{eqnarray}}

\newcommand{\dd}{{\delta}}

\newcommand{\no}{\noindent}

\newcommand{\bc}{\begin{center}}
\newcommand{\ec}{\end{center}}

\newcommand{\fref}[1]{Figure\ \ref{fig:#1}}

\newcommand{\tref}[1]{Table\ \ref{tab:#1}}

\newcommand{\eref}[1]{eqn.\ (\ref{eqn:#1})}
\newcommand{\erefs}[2]{eqns.\ (\ref{eqn:#1})--(\ref{eqn:#2}\mbox{)}}

\newcommand{\id}{\mathrm{d}}
\newcommand{\Dt}[1]{\frac{\id #1}{\id t}}

\newcommand{\mat}{\begin{array}{cc}}
\newcommand{\emat}{\end{array}}

\newcommand{\LD}{\LD_{50}}